\documentclass[%
 aps,
 prd,
 superscriptaddress,
 amsmath,amssymb,
 aps,nofootinbib,   
]{revtex4-2}

\usepackage[a4paper,left=1.5cm,right=1.5cm,top=3cm,bottom=3cm]{geometry}

\usepackage{bm}
\PassOptionsToPackage{linktocpage}{hyperref}
\usepackage[hyperindex,breaklinks]{hyperref}
\usepackage{enumitem}
\usepackage{slashed}

\usepackage{float}
\usepackage{yfonts}

\usepackage{amsfonts}
\usepackage[dvipsnames]{xcolor}
\usepackage{graphicx}
\usepackage{longtable}
\usepackage{verbatim}
\usepackage{color}
\usepackage{mdframed}
\usepackage{soul}
\usepackage{amsfonts,amssymb,mathrsfs,amsmath,esint}
\usepackage{slashed, cancel}
\usepackage{framed}
\usepackage{mdframed}
\usepackage{simplewick} 
\allowdisplaybreaks

\usepackage{latexsym}
\usepackage{graphicx}
\graphicspath{{./Figures/}}
\usepackage[dvipsnames]{xcolor}
\usepackage{booktabs}
\usepackage{datetime}
\newdateformat{mydate}{\THEDAY{ }\monthname[\THEMONTH]{ }\THEYEAR}

\usepackage{tikz}
\usepackage{color}
\usepackage{framed}
\usepackage{hyperref}
\hypersetup{colorlinks, citecolor=bluscuro, linkcolor=black, urlcolor=bluscuro}
\definecolor{rossos}{cmyk}{0,1,1,0.55}
\definecolor{bluscuro}{rgb}{0.15, 0.2, .85}
\definecolor{bluchiaro}{cmyk}{1,.3,0.,0.1}

\newcommand{\be}{\begin{equation}}
\newcommand{\ee}{\end{equation}}
\newcommand{\ba}{\begin{eqnarray}}
\newcommand{\ea}{\end{eqnarray}}

\newcommand{\E}{\mathcal{E}}

\newcommand{\e}{\rho}

\begin{document}

\title{The quantum de Sitter root of quasi de Sitter observables}
 
\author{Cesar Gomez} 
\affiliation{Instituto de F\'{i}sica Te\'orica UAM-CSIC, Universidad Aut\'onoma de Madrid, Cantoblanco, 28049 Madrid, Spain.}
\email{cesar.gomez@uam.es}

\author{Raul Jimenez} 
\affiliation{ICC, University of Barcelona, Marti i Franques 1, 08028 Barcelona, Spain.}
\affiliation{ICREA, Pg. Lluis Companys 23, Barcelona, E-08010, Spain.}
\email{raul.jimenez@icc.ub.edu}


\begin{abstract}
In inflationary cosmology the quasi de Sitter graceful exit allows us to measure the quantum features of the primordial dS phase, in particular, the lack of scale invariance parametrized by the spectral index $n_s$. In this article we summarize previous work on how the underlying primordial scaling law is implemented in the dS quantum Fisher information of the dS planar ground state (dSQFI). At large scales the dSQFI unequivocally sets, without any qdS input, the value of $n_s$ to be 0.9672. This value is independent of the tensor to scalar ratio whose value requires model dependent input. In addition the dSQFI predicts, at large scales, a small running compatible with the current experimental results. Other phenomenological consequences of the dSQFI for small scales, will be discussed in a future publication. 
\end{abstract}

\maketitle

\section{introduction}
\begin{figure}
    \centering
    \includegraphics[width=0.7\columnwidth]{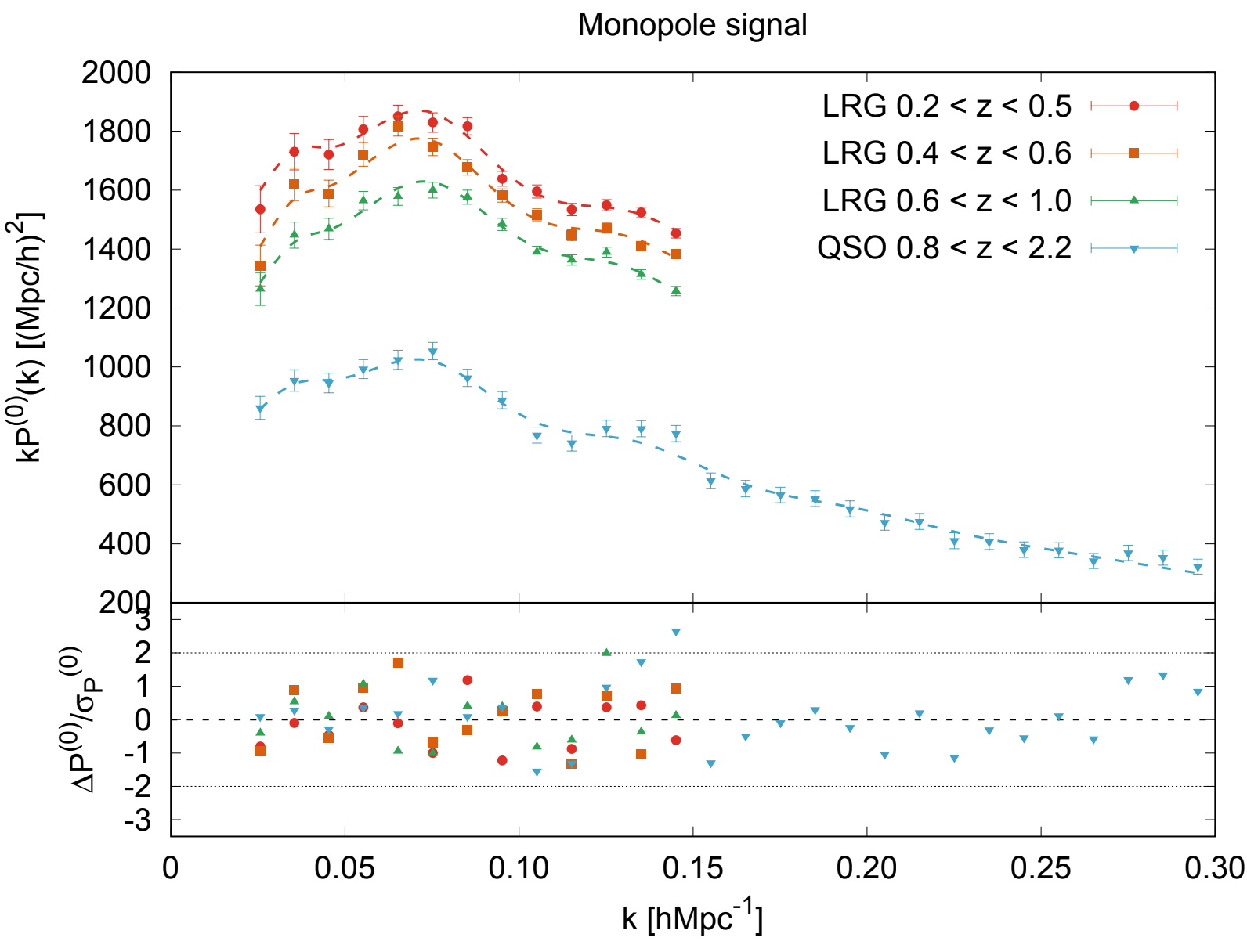}
    \caption{The power spectrum of galaxies and QSOs as derived from the latest LSS data from the BOSS and eBOSS surveys. (Hector Gil-Marin, private communication)}
    \label{fig:BOSS}
\end{figure}

One of the greatest achievements of inflationary Cosmology is to encapsulate the early source of the Universe present complexity into a pure number: the spectral index. This number measures the anomalous scale invariance of the power spectrum of primordial scalar quantum curvature fluctuations. Quantum effects in an early phase of exponential expansion lead to curvature fluctuations that are adiabatic, almost gaussian  and very close to scale invariant. Once a graceful exit of the expanding phase is implemented, the primordial anomalous scale invariance is enhanced by gravitational instability delivering the observed large scale structure of the present Universe. This structure can be measured using Cosmic Microwave Background (CMB) and Large Scale Structure (LSS) data setting the experimental value of the spectral index with high precision (see Fig.~\ref{fig:BOSS} for the last available data from LSS using galaxies and QSOs). 

In the present inflationary paradigm, the value of $1-n_s$ depends on phenomenological input: the particular quasi-de-Sitter model used to define an early expanding phase with a graceful exit. Generically, these models are fully determined by the potential energy of the inflaton field with the different gradients of the potential defining the {\it slow-roll} signature of the quasi de Sitter phase. The key information on the curvature quantum fluctuations is encoded in the corresponding power spectrum $\mathcal{P}$. This power spectrum satisfies a Renormalization-Group (RG) equation
\begin{equation}
    k \frac{d}{d k} \mathcal{P} = (1-n_s) \mathcal{P}
    \label{eq:RG}
\end{equation}
where the spectral index $(1-n_s)$ plays the role of a formal {\it beta function} with scale invariance corresponding to $n_s=1$, a case known as the Harrison Zeldovich spectrum.

This analogy with the RG motivates a natural question, namely, the {\it universality} of the spectral index. Our experience with the physics of critical phenomena, as described using the Wilsonian renormalization group, teaches us that around scale invariant critical points the departures from criticality are determined by scaling laws of the type $(T-T_c)^{\gamma}$ with the critical exponent $\gamma$ enjoying a high degree of universality only dependent on {\it symmetries} and {\it dimensionality}. This universality of the exponents contrasts with the microscopic dependence of the particular value of, for instance, the critical temperature $T_c$. Thus it looks natural to ask ourselves :\\
Is $1-n_s$ the analog of a "critical exponent" for a ``de Sitter Universality Class'' only depending on space-time dimension and the {\it quantum} implementation of de Sitter symmetries? Or perhaps in more concrete terms: Is equation (\ref{eq:RG}) describing the anomalous quantum implementation of scale transformations in de Sitter?

The answer to this question will be yes. Moreover, we will show how this approach predicts unequivocally the value of $1-n_s$ in a fully model independent way deeply rooted in the renormalization group equation. The rest of this article, based on recent lectures, will intend to provide, in a pedagogical manner,  the basic technical tools used in \cite{GJ4,GJ5,GJ6,GJ7} to demonstrate this statement \footnote{For previous work on the potential use of quantum information in Cosmology see \cite{GJ1,GJ2,GJ3}.}.

\section{Brief Review on Inflation}

Historically, we can identify two basic initial approaches to Inflation. One approach,  initiated by Starobinsky, Mukhanov, Chivisov and others \cite{Inf0,Inf1, Inf2,MCH} was motivated by the initial singularity problem and was based on the study of the General Relativity consequences of the contribution of the vacuum polarization effects to the trace of the energy momentum tensor. The other approach, initiated by Guth \cite{Inf3} and further developed by Linde, Albrecht and Steinhardt  \cite{Inf4,Linde,Albrecht}  was concerned with the horizon and flatness problem. In this approach, the key conceptual ingredient was the notion of false vacuum and bubble supercooling. The merging of these two approaches happened very soon only a  few years later~\cite{Kofman:1985aw}. Since there are many excellent reviews of inflation we will focus on this section on some concrete comments relevant for the future discussion on the spectral index\footnote{The other definitive prediction of inflation is the value of non-gaussianities; in single field inflation this is equal, in the squeezed limit, to precisely $1-n_s$. For some recent works on the observability of this feature see e.g.~\cite{GJ3,Bertacca} }. We first focus on the model independent predictions of inflation.

\subsection{Model Independent Considerations}

\subsubsection{Curvature Fluctuations}

We briefly review the power spectrum of scalar curvature fluctuations as originally derived in~\cite{MCH}. This derivation was done in the framework of the first Starobinsky model characterized by the parameters $H$ and $M$ as defined by the trace anomaly. Irrespectively of the trace anomaly meaning of $M$ as defined by the value of $k_3$ let us take $M$ as a free phenomenological parameter, this ensures that it will be model independent in the sense that the value of $M$ will be later fixed by observations. In the large scale regime we can use the Sachs-Wolfe~\cite{Sachs:1967er} effect to relate the temperature fluctuations of the CMB $\frac{\delta T}{T}$ along a given direction $e$ and location $x_0$ with the curvature fluctuations $\chi$ through the relation
\begin{equation}
    \frac{\delta T}{T} \sim \frac{1}{5} \chi(\eta_e- e(\eta_e-\eta_0)+x_0)
\end{equation}
with $\eta_e$ the emission conformal time and $\eta_0$ the present conformal time. The curvature fluctuation $\chi$ is defined as
\begin{equation}
    \chi = \frac{\Phi}{a\sqrt{2\epsilon} M_P}
\end{equation}
with $\epsilon$ the first slow roll parameter and $\Phi$ satisfying \cite{CHM}

\begin{equation}\label{key}
    \Phi^{,,}_k + \left (k^2 - \frac{z^{,,}}{z} \right ) \Phi_k =0
\end{equation}
with $z= a \sqrt{\epsilon}$. Let us now define the simplest one parameter family of slow roll models characterized by
\begin{equation}\label{zeta}
     \frac{z^{,,}}{z} = \frac{\beta (\beta + 1)}{\eta^2}
\end{equation}
with $\beta =-2 - \delta$. Solving (\ref{key})
for the case $|k \eta| << 1$ and after imposing, as initial condition, $\Phi_k = \frac{1}{\sqrt{2k}} e^{ik\eta}$ for $|k\eta| >> 1$ yields
\begin{equation}\label{ONE}
    k^3 | \Phi_k (k \eta) |^2 \sim \frac{1}{\eta^2} (k \eta)^{-2 \delta}
\end{equation}
The power spectrum for scalar perturbations is now defined as follows. For a given momentum $k$ we define the "horizon exit" time by the condition $k\eta =1$. The amplitude of the power spectrum $\mathcal{P_{\chi}}(k)$ at this time is given by
\begin{equation}
    \mathcal{P_{\chi}}(k) =  \frac{H^2}{\epsilon M^2_p}
\end{equation}
with $H$ and $\epsilon$ evaluated at the time the mode of momentum $k$ exits the horizon. Let us now set a {\it pivot time scale} $\eta_0= \frac{1}{k_0}$. The power spectrum for $k$ close to the pivot scale $k_0$ is then given using (\ref{ONE}) as 
\begin{equation}\label{TWO}
    \mathcal{P_{\chi}}(k;k_0) =  \frac{H_0^2}{\epsilon_0 M^2_p}(\frac{k_0}{k})^{2\delta} \equiv \mathcal{P_{\chi}}(k_0)(\frac{k_0}{k})^{2\delta}
\end{equation}
with $H_0$ and $\epsilon_0$ the values of $H$ and $\epsilon$ at the pivot time scale, namely at the time the pivot momentum $k_0$ exits the horizon. The spectral index $(1-n_s)$ at this pivot scale is given by
\begin{equation}
    1-n_s = 2 \delta
\end{equation}
The renormalization group meaning of (\ref{TWO}) is now pretty obvious. If we take the pivot scale as the analog of the renormalization point and we define the RG as the independence on the pivot scale we get
\begin{equation}
    k_0 \frac{d\mathcal{P_{\chi}}(k_0)}{dk_0} = 2\delta \mathcal{P_{\chi}}(k_0)
\end{equation}
that leads to the well known relation between the spectral index and the slow roll parameters
\begin{equation}\label{qdS}
    \delta = 3\epsilon - \eta
\end{equation}
with $\epsilon$ and $\eta$ the first and second slow-roll parameters evaluated at the pivot scale. In summary the Schrodinger equation (\ref{key}) for the metric (\ref{zeta})  sets the value of $\delta$ and the RG equation determines the relation between $\delta$ and the slow roll parameters $\epsilon$ and $\eta$.

\subsubsection{Quasi de Sitter metric}
Let us now repeat the former exercise introducing the notion of quasi de Sitter metric formally defined by the equation:
\begin{equation}\label{metric}
    \frac{a^{,,}_{qdS}}{a_{qdS}} \equiv \frac{z^{,,}}{z} = \frac{\beta (\beta + 1)}{\eta^2}
\end{equation}
with again $\beta =-2 - \delta$ \footnote{ See \cite{MV1,MV2} and references therein.} Note that $a_{qsS} = a \sqrt{\epsilon}$ however in (\ref{metric}) we only introduce as input the value of $\delta$. The solution to (\ref{metric}) can be written as
\begin{equation}\label{metric2}
    a_{qdS}= \frac{-1}{H_0 \eta} \frac{1}{(k_0 \eta)^{\delta}}\sqrt{\epsilon_0}
\end{equation}
where at this level $H_0$ and $\sqrt{\epsilon_0}$ are simply two integration constants with the only difference that $H_0$ has dimensions of energy while $\epsilon_0$ is dimensionless and $k_0$ is a pivot scale. 
The power spectrum is now simply defined as
\begin{equation}
    \mathcal{P_{\chi}}(k;k_0) =\frac{1}{M_P^2 a_{qdS}^2}k^3 | \Phi_k (k \eta) |^2 
\end{equation}
that leads to the former relation, namely
\begin{equation}
    \mathcal{P_{\chi}}(k;k_0)= \frac{H_0^2}{\epsilon_0 M_P^2}(\frac{k_0}{k})^{2\delta}
\end{equation}
with the integration constants $H_0$ and $\epsilon_0$ representing the values of $H$ and $\epsilon$ at the pivot scale $k_0$.

In the regime of large scales where we use the Sachs-Wolfe relation to define the observed spectrum of energy fluctuations, the pivot scale $k_{CMB}$ exits the horizon before the end of inflation. The key property supporting inflation is that for these scales the quantum fluctuations leaving the horizon before recombination behave {\it coherently} \cite{Coulson,Dodelson}; a necessary condition to create the acoustic CMB picks. Moreover, in this regime, the variation of the value of the slow roll parameters is very small and therefore we can locally {\it parametrize} the CMB region using a quasi de Sitter metric of the type described above. Thus, we can describe the large scale fluctuations around the CMB scale with a quasi de Sitter defined by a $\delta_{CMB}$ with $(1-n_s)_{CMB} = 2\delta_{CMB}$.

In this sense the key question we want to address is:

{\it Can we predict the value of $\delta_{CMB}$ using only pure quantum de Sitter information?}

Although the rest of this article will be dedicated to review the way to answer technically this question, it could be useful to provide some intuition first. As stated in the previous paragraphs, the spectral index contains crucial {\it information} about the observed spectrum after reentering. In this sense, it encodes the information about the primordial spectrum that is safely transferred to the reentering moment and therefore observable. For that to work, the coherence of phases is, as already mentioned, a necessary ingredient. On the basis of this intuition we can try to extract the maximal quantum information we can have in pure de Sitter on what, for a local observer, exits the horizon. Using the notion of quantum Fisher information, to be discussed later, we are able to extract this information. Imposing that this information is, for large scales, precisely the one we observe at reentering time uniquely sets the value of $\delta_{CMB}$.

\subsection{Model Dependent Considerations}

\subsubsection{Vacuum polarization and Trace Anomaly}
It is due to Starobinsky~\cite{Inf1} the key observation that once the effects of vacuum polarization for conformal matter are included in the Einstein equations, there exists a solution of de Sitter type. The effect of vacuum polarization for conformal matter is fully encoded in the trace anomaly (see~\cite{Duff:1993wm} for references and a nice review):
\begin{equation}
<T^{\mu}_{\mu}> = - \frac{1}{2880 \pi^2} \left ( k_1 C_{i \alpha l m} C^{i \alpha l m} + k_2 \left ( R_{i \alpha} R^{i \alpha} - \frac{1}{3} R^2 \right ) + k_3 \square R \right )
\end{equation}
with the coefficients $k_1,k_2,k_3$ being determined by the matter content. If we assume a conformally flat space-time geometry $ds^2 = -dt^2 +a^2(t) dx^2$, then 
 Einstein field equations
\begin{equation}
R_{ik}-\frac{1}{2}g_{ik} R = 8 \pi G < T_{i k} >    
\end{equation}
have a de Sitter solution with $a= \frac{-1}{\eta H}$, $\eta$ the conformal time and with Hubble parameter $H^2 = \frac{360 \pi}{G k_2}$ for $G$ the Newton constant. In addition, the piece of the trace anomaly with coefficient $k_3$ provides the needed information to analyze the stability of this particular solution. Indeed, this component of the trace anomaly leads to the existence of the so called {\it scalaron} with mass given by $M^2 = \frac{-360 \pi}{G k_3}$. Roughly speaking, the quantum life time of this {\it scalaron} gives us the expected stability time of this particular de Sitter solution. In this sense, the quantum effects encoded in the trace anomaly seem to provide the needed ingredients to give rise to a primordial de Sitter phase with a graceful exit set by the value of the scalaron mass. In the original approach by Starobinsky, the existence of a de Sitter solution was welcomed as a natural solution to the initial singularity problem and not as the current inflationary mechanism to solve the horizon and flatness problem. This beautiful picture is, however, too much constrained by the particular spectrum of conformal matter contributing to the trace anomaly. For the model to work, $k_3$ must be negative and $|k_3| >> k_2$ something that is generically difficult to achieve in realistic models (see for a more recent discussion \cite{Hawking}). 

This difficulty was surpassed by the observation that the $k_3$ term in the trace anomaly can be re-absorved by adding a local term
\begin{equation}
    \Delta \mathcal{L} = \frac{M^2_p R^2}{96 \pi M^2}
\end{equation}
to the Einstein Lagrangian~\cite{Kofman:1985aw}. The so defined $R^2$ gravity propagates, in addition to the spin two graviton, a scalar field that plays the role of the inflaton field for the, now known, as the Starobinsky model \footnote{The inflaton potential for this model is given by:
$V(\phi)=\frac{3}{4} M_P^2M^2( 1- e^{\sqrt{\frac{2}{3}}\phi/M_P})^2$. Where as usual $M_P$ is the Planck mass and $\phi$ is the canonically normalized field \cite{Maeda} ( see \cite{Luis} for a recent discussion on $R^2$ gravity).}.

\subsubsection{Tensor fluctuations}
Before ending let us say few words about tensor fluctuations and the ratio $r$ of tensor to scalar.The first thing to be noticed is that the Schrodinger equation for each polarization of the tensor fluctuations is the same as for the scalar fluctuations namely (\ref{key}). Up to numerical factors irrelevant for the present general discussion the power spectrum for tensor modes is given by
\begin{equation}
\mathcal{P}_{t}(k;k_0)= \epsilon(k;k_0)\mathcal{P_{\chi}}(k;k_0)
\end{equation}
with
\begin{equation}
    \epsilon(k;k_0) = \epsilon_0 (\frac{k_0}{k})^{\Delta}
\end{equation}
with $\Delta= 2\eta$ ($\eta$ the second slow-roll parameter) being again determined by the renormalization group invariance of $\mathcal{P}_t(k;k_0)$ with respect to changes in the pivot scale. This is the well known result that sets both the tensor to scalar ratio $r$ as well as the tensor spectral index to be $O(\epsilon)$.

Note that neither $r$ nor the tensor spectral index is uniquely determined by the value of $\delta$. This will be important for our future discussion. 

\subsubsection{Tilt and de Sitter time scales}

Regarding the relevant time scales of de Sitter is important to distinguish three different cases:
\begin{itemize}
    \item The time scale of quantum instability of pure dS.
    \item The time scale on which an initial quasi de Sitter metric enters into a Friedman expansion phase.
    \item The number of e-foldings between the moment the scale $k_{\rm CMB}$ exits the horizon and the end of inflation.
\end{itemize}
Concerning quantum instability of pure de Sitter this scale depends on the quantum model of de Sitter space time as a coherent state of gravitons defined relative to a fundamental Minkowski space \cite{DG,D,Bra2}. This scale is known as {\it quantum break time} and it is given by
\begin{equation}
    T_{qb} = \frac{M_P^2}{H^3}
\end{equation}
A different approach to define an intrinsic quantum instability of pure de Sitter is based on defining for de Sitter the analog of Unruh vacuum \cite{Aalsma,Verlinde} instead of the eternal Bunch Davis vacuum. 

Moreover one can define a different quantum time scale based on what we can denote as {\it de Sitter anomalies} \cite{Dyson,susskind}. Generically these anomalies are defined by the variance of de Sitter generators that are not closed on the static patch of a given observer. These variances are non zero due to the discretenes of the spectrum of the static patch Hamiltonian. These anomalies lead to typical recurence times. Finally, another quantum time scale that can be defined {\it ab initio} is the one based on the so called trans-Planckian censorship \cite{vafa} (see \cite{Bran} for a recent review  and \cite{Ulf2}). In this case, the time scale is set by imposing that transplanckian modes cannot exit the de Sitter horizon. 

\begin{figure}
    \centering
    \includegraphics[width=0.5\columnwidth]{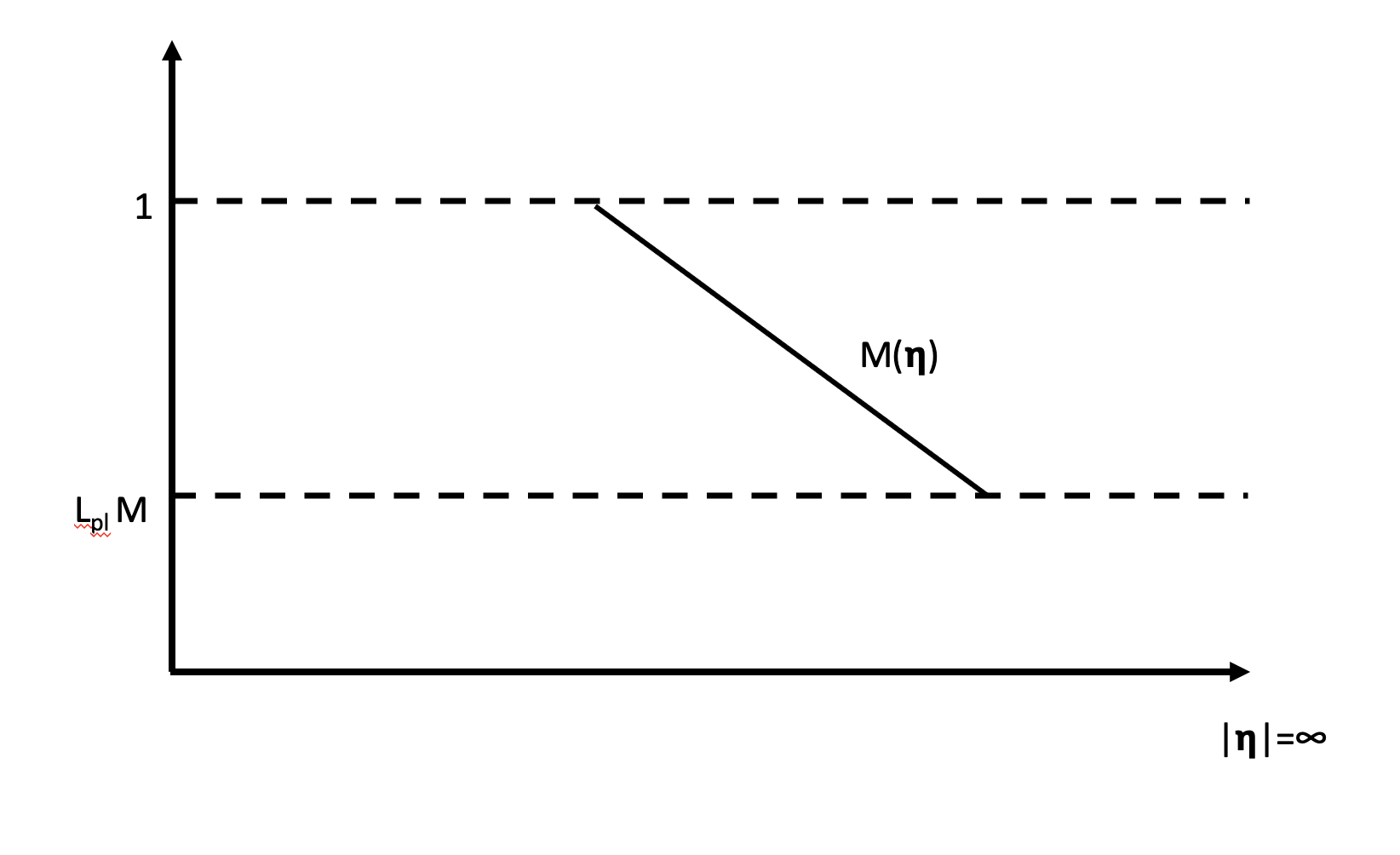}
    \caption{Time scale in inflationary models. The length of the inflationary patch $L_{\rm pl}$ is plotted as a function of increasing time $|\eta|$.}
    \label{fig:Fig1}
\end{figure}

With respect to the typical time scale on which an initial quasi de Sitter enters into a Friedmann expanding phase the original discussion is contained in~\cite{MCH}. The simplest way to get this time scale (see Figure 1 in~\cite{MCH} and Fig.~\ref{fig:Fig1}) is using the tilt of the power spectrum to define an effective scalaron mass as
\begin{equation}
    M^2(\eta) = M^2 (k_0\eta)^{(1-n_s)}
\end{equation}
and to define the time  scale by $M^2(\eta)\sim M_P^2$. This leads to
\begin{equation}
    t_{qdS}= \frac{1}{H}\frac{1}{(1-n_s)}\ln(\frac{M_P^2}{M^2})
    \end{equation}
    with $(1-n_s)= \frac{2M^2}{3H^2}$.

Finally, we have the time scale associated with the number of e-foldings $\mathcal{N}_{\rm CMB}$ measuring the time  between the moment the CMB scale $k_{CMB}$ exits the horizon and the end of inflation. This quantity is expected to be between 50 and 60 for our Universe. As already pointed out in \cite{MCH} in the Starobinsky model this quantity is given in terms of the tilt by the relation
\begin{equation}
    (1-n_s)_{\rm CMB}\sim \frac{2}{\mathcal{N}_{\rm CMB}}+ O(\frac{1}{\mathcal{N}_{\rm CMB}^2})
\end{equation}
As discussed above our target will be to discover the quasi de Sitter parametrization $2\delta_{CMB} = (1-n_s)(CMB)$. Note that $\delta_{CMB}$ is related to the Starobinsky slow roll parameters through relation (\ref{qdS}).

\section{de Sitter Geometry and Vacuum}

\begin{figure}
    \centering
    \includegraphics[width=0.4\columnwidth]{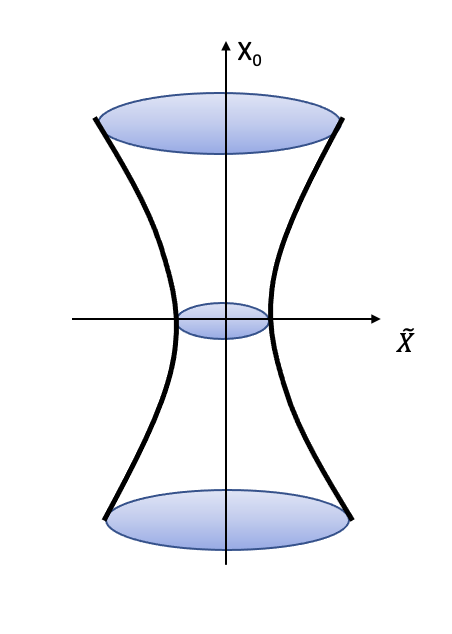}
    \caption{de Sitter space-time metric}
    \label{fig:dS}
\end{figure}

de Sitter space-time in $d$ dimensions is defined as the hyperboloid (see Fig.~\ref{fig:dS}) 
\begin{equation}
    (dS)_{\alpha} = -X_0^2 + X_1^2+...+X_d^2 = l^2
\end{equation}
in $d+1$ Minkowski with coordinates
 $(X_0...X_d)$. We can define global coordinates as 
\begin{equation}
  ds^2 = -d\eta^2 + \cosh^2 (\tau) d\Omega_{d-1} = \frac{1}{\cos^2 T} \left ( -d T^2 + d \Omega^2 \right )_{-\pi/2 < T < \pi/2}  
\end{equation}

\begin{figure}
    \centering
    \includegraphics[width=0.4\columnwidth]{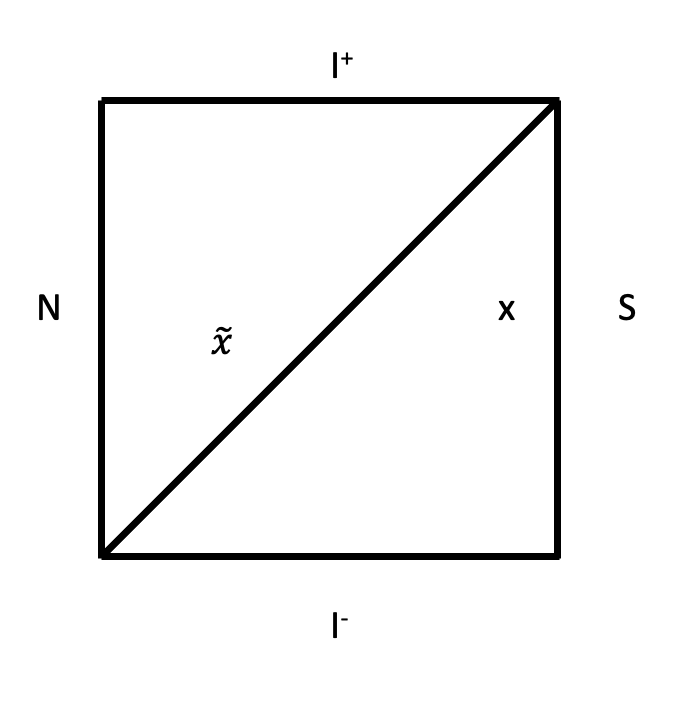}
    \caption{The causal past of S-observer and the casual future of N-observer. Points $x$ and $\Tilde{x}$ are related by an antipodal mapping (see text for further details).}
    \label{fig:Penrose}
\end{figure}

\begin{figure}
    \centering
    \includegraphics[width=0.4\columnwidth]{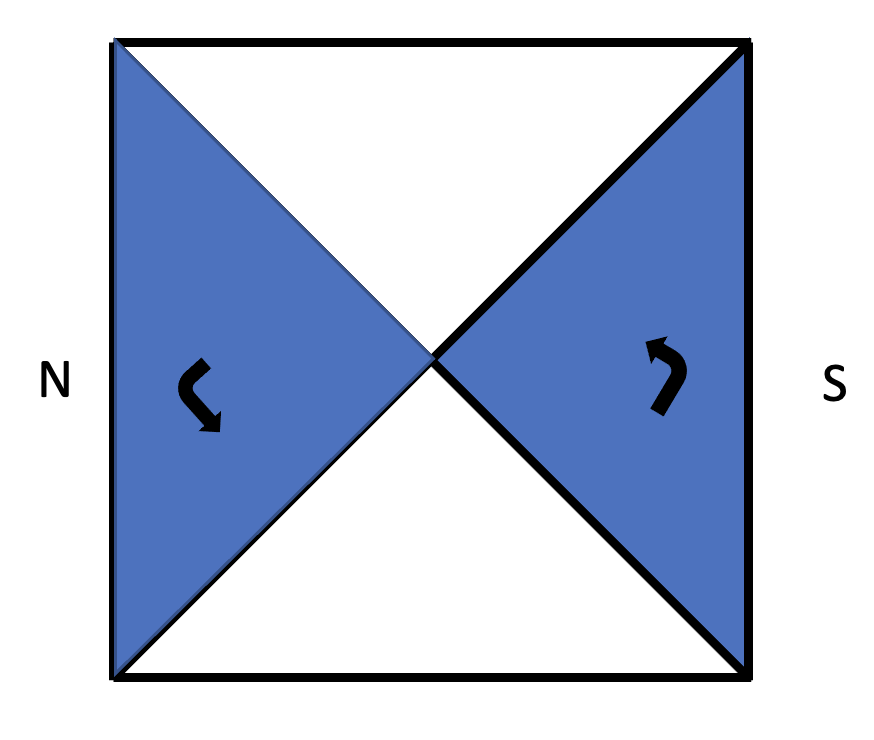}
    \caption{North and South static patches. The arrows depict the time flow in the Penrose diagram of de Sitter.}
    \label{fig:Fig4}
\end{figure}

The Penrose diagram associated to this metric is given in Fig.~\ref{fig:Penrose}. We can consider two different observers located at the north and south pole. The causal past and causal future of these observers are the two triangular patches in the figure. An important property  of classical de Sitter space time can be directly extracted from its Penrose diagram. Namely, in order to define asymptotic data on both $I^+$ and $I^-$ we need the two observers that are causally disconnected. This creates the classic problem on the standard definition of a de Sitter S-matrix. Another interesting aspect is the existence of isometries in the de Sitter group that are connecting the causal future and causal past of the two observers. This will have an immediate consequence
once we try to define quantum field theory on dS (see~\cite{Schr,ver}). Namely,  quantum states invariant under these transformations will be associated with quantum {\it entanglement} between the north and the south observer.

A transformation that will play an important role in the definition of de Sitter invariant vacua is the  anti-podal map:
\begin{equation}
   X=( X_0,X_i) \rightarrow \tilde X=(-X_0,-X_i)
\end{equation}
The geodesic distance in dS, between two points with Minkowski coordinates $x$ and $x'$ is defined by $\cos \mathcal{D}(x,x^{'}) \equiv P(x, x^{'}) = \eta_{ij}x^i x^j$. From this we observe that points separated by null intervals correspond to $P(x,x^{'})=1$. Moreover since $P(x,x^{'})=-P(x,\Bar{x^{'}})$ for two points related by the antipodal map we get $P(x,\tilde{X})=-1$. 

\subsection{de Sitter invariant vacua}
Next, we will be interested in defining a quantum field theory (QFT) in dS. Let us consider the case of a free scalar field $\hat\Phi$. The QFT will be defined by the Wightman function $W(x,x')= \langle \hat\Phi(x)\hat\Phi(x')\rangle$. We should expect that this Wightman function depends on the geodesic distance between $x$ and $x'$ i.e. to be a function of $P(x,x')$ analytic except for points separated by null intervals i.e. for $P=1$. The equation of motion satisfied by $W(P)$ for a free scalar field of mass $m$ is given by \cite{Spradlin,Bousso}
\begin{equation}
(1-P^2)\partial_P^2 W -dP\partial_PW -m^2W =0
\end{equation}
Since this equation is invariant under the antipodal transformation $P\rightarrow -P$ if $W(P)$ is a solution then $W(-P)$ is also a solution and consequently also any linear combination of both.
Using the Euclidean continuation of de Sitter we get the Euclidean solution $W_E(P)$. For the corresponding mode expansion
\begin{equation}
    \hat{\Phi} (x) = a_{n}^{E} \Phi_{n}^{E} (x) + a_{n}^{E \dagger} \Phi_n^{E *} (x)
\end{equation}
we define the Heisenberg algebra
\begin{equation}
    [a_{n}^{E},a_{m}^{E \dagger}] = \delta_{n,m}
\end{equation}
and the Euclidean vacuum $a_{n}^{E} |0> =0$ in such a way that $W_E(x,x') = \langle0|\hat \Phi(x) \hat \Phi(x')|0\rangle$. Choosing the euclidean modes satisfying $\Phi_n(x)^* = \Phi_n(\tilde x)$ we can define the family of $\alpha$ vacua \cite{CT, Tagirov,vacua1,vacua2,Banks, Einhorn, Ulf} using the linear superpositions of Wightman functions
\begin{equation}
    W_{\alpha}(x,x') = C^2(\alpha) ( W_E(x,x') + e^{\alpha+\alpha^*} W_E(x',x) + e^{\alpha}W_E(x,\tilde x') + e^{\alpha^*}W_E(\tilde x,x'))
\end{equation}
With $C^2(\alpha) = \frac{1}{1-e^{\alpha+\alpha^*}}$. The $\alpha$ vacua $|\alpha\rangle$ associated with this family of Wightman functions is given by
\begin{equation}\label{vacua}
    |\alpha> = C e^{\{ \sum_n \frac{1}{2} e^{\alpha *} (a_{n}^{E \dagger})^2 \} } |0>
\end{equation}
These are de Sitter invariant vacua with the Euclidean vacua corresponding to ${\rm Re}(\alpha)=-\infty$. Note also that for non vanishing $Im(\alpha)$ the corresponding vacuum breaks CPT. There exist an extensive discussion in the literature on the physical meaning of these vacua, in particular if we consider a self-interacting scalar field (see \cite{Banks} and \cite{Einhorn}). We will not enter into this discussion. An important aspect of the so defined $\alpha$ vacua is that the exponent in (\ref{vacua}) is independent of the mode $n$.

\subsection{Quantum Fisher information about the $\alpha$ parameter.}
The introduction of de Sitter invariant $\alpha-$vacua provides a good excuse to introduce some basic notions on Quantum Fisher Information (QFI)[for a review see~\cite{Paris} and Appendix~\ref{app:fisher}]. Irrespectively of the physical meaning of the $\alpha$ parameter, the $\alpha-$vacua are a good example of a family of quantum states described by a parameter that is not associated with any self-adjoint observable. In other words, we don't have any observable whose eigenvalues can be associated with the actual value of the parameter $\alpha$. In these conditions a fair question is: {\it How much information can we have about the actual value of $\alpha$ using as input the results of an arbitrary set of measurements for self-adjoint observables?} The upper bound on this information is what defines the quantum Fisher information the quantum Fisher information about this parameter \footnote{A natural example of a similar phenomana is the $\theta$ parameter in QCD. In this case it was suggested in \cite{gomez} that the quantum Fisher information is determined by the topological susceptibility.}. In the case of de Sitter invariant $\alpha-$vacua this question is purely academic and formal. The reason is that the de Sitter invariant vacua are globally defined and the corresponding information cannot be associated with one observer. Thus this formal quantum Fisher will be a sort of meta-information in a similar way that meta-parameters are used in hierarchical Bayesian estimation theory. In any case, although we will not use this particular global quantum Fisher information, we think it is a good place to introduce the notion of quantum Fisher. 

In order to do that let us first introduce the generator associated with changes of $\alpha$, namely
the $\alpha-$Hamiltonian
\begin{equation}
    \frac{\partial}{\partial \alpha_{I}} |\alpha>  =  H^{I} |\alpha > 
    \end{equation}
    that yields
    \begin{equation}
    H^{I}  =  \frac{-i}{2} \sum_{n} (a_{n}^{E \dagger} )^{2} e^{\alpha^{*}} + c.c.
\end{equation}
where we only care about changes in the imaginary part of $\alpha$. The QFI for the $\alpha$ parameter is simply defined as the variance of $H^{I}$ i.e.
\begin{equation}\label{fisher}
    \mathcal{F}(\alpha) = <\alpha| (H^{I})^2 | \alpha > - \left ( < \alpha | H^{I} | \alpha > \right )^2
\end{equation}
This formal quantity measures, as already pointed out, the maximum information we can get on the value of $\alpha$ performing measurements (and meta-measurements) on the $|\alpha>$ vacua. More precisely for the {\it quantum estimator} of $\alpha$ let us call it $O_{\alpha}$ \footnote{This is the self adjoint operator formally playing the role of the conjugate of $H^{I}$.}  we have
\begin{equation}
    \langle \Delta O_{\alpha}^2\rangle = \frac{1}{\mathcal{F}(\alpha)}
\end{equation}
Note that this is not equal to the standard statistical variance defined as $\frac{1}{M\mathcal{F}(\alpha)}$
for $M$ the number of actual measurements.

In what follows we will define a modified notion of the QFI that accounts for the actual information accessible to a physical observer. Before doing this it is worth to briefly discuss the interplay between de Sitter invariance and entanglement.

\subsection{de Sitter invariant vacua and quantum entanglement}
Let us now consider the static patches for the north and south observers as depicted in Fig.~\ref{fig:Fig4}. On these patches we have globally defined time as shown by the arrows in the figure. Let us formally define for the North and South observers the corresponding Hamiltonians and assume a discrete energy spectrum $E_i^{N},E_i^{S}$. A globally de Sitter invariant $|\alpha\rangle$ vacua will correspond to a particular quantum state, living in the tensor product of the two static Hilbert spaces, of the type
\begin{equation}
\sum C_{i,j}(\alpha) |E_i^{N}\rangle |E_j^{S}\rangle
\end{equation}
Tracing over the North will define a density matrix for the South observer with the corresponding von Neuman entropy.

The particular euclidean vacua is the TFD (thermo-field double) state and the corresponding density matrix, after tracing over one of the two static observers is just the GH (Gibbons-Hawking) thermal density matrix with vN entropy equal to the well known GH de Sitter entropy \cite{GH}. As shown in the Penrose diagram this euclidean $|\alpha\rangle$ vacua describes the eternal AdS black hole. 

To induce any form of quantum instability i.e. evaporation of this black hole requires to {\it break some de Sitter symmetries} as it is the case with Unruh like vacua ( recall discussion in section C).

Moreover, for the static observer we can easily introduce the notion of de Sitter anomalies. Take any generator $G$ of the de Sitter group that moves points in one of the static patches, for instance the South patch, out of the patch. The {\it anomaly} will be simply defined by the variance of this generator relative to the energy ground state chosen by the south observer. We will not enter into this subtle discussion, the interested reader can see~\cite{susskind} and references therein \footnote{A different form of entanglement entropy associated with tracing over modes in the planar patch was discussed in \cite{Bran2,Bran3,Br}.}.

\section{Inflation in the planar patch}

\begin{figure}
    \centering
    \includegraphics[width=0.5\columnwidth]{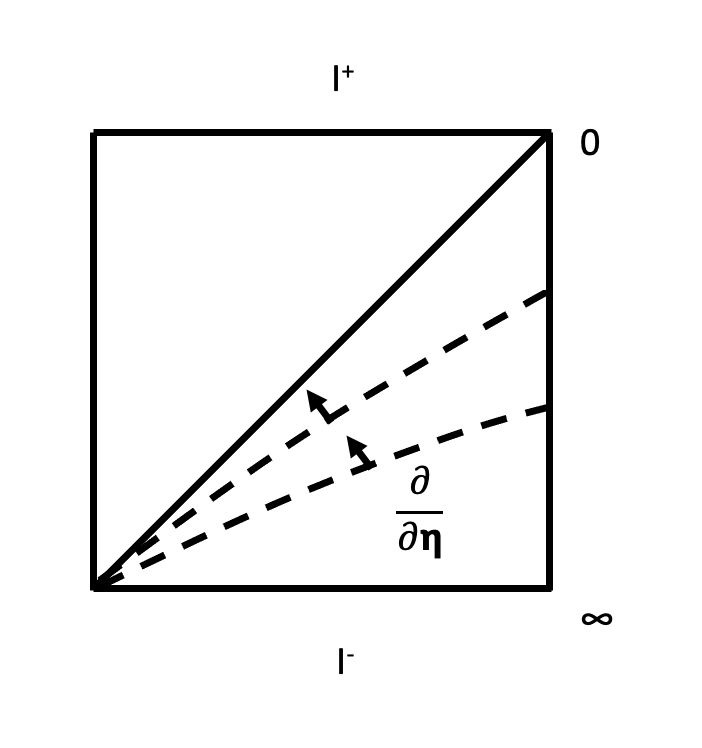}
    \caption{Hypersurfaces of constant $\eta$}
    \label{fig:Fig5}
\end{figure}

\begin{figure}
    \centering
    \includegraphics[width=\columnwidth]{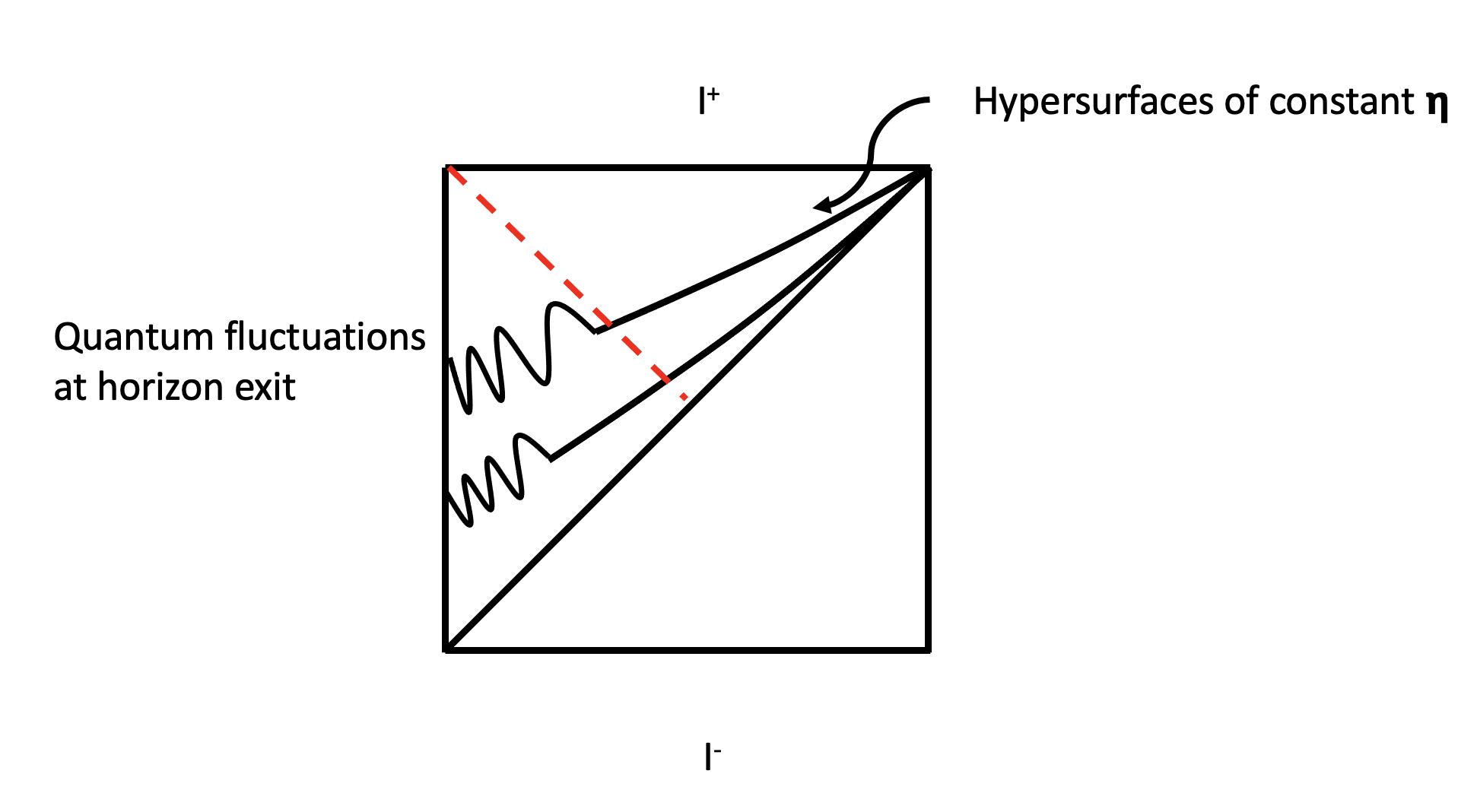}
    \caption{The evolution of quantum fluctuations.}
    \label{fig:Fig6}
\end{figure}

The de Sitter metric describing an expanding Universe is defined using planar coordinates 
\begin{equation}
ds^2 = -dt^2 + a(t)^2 dx^2 = \frac{1}{H^2 \eta^2} \left (-d \eta^2 + dx^2 \right )
\end{equation}
where the conformal time $\eta = \frac{-1}{aH}$. These coordinates are only covering what we can denote the planar patch that corresponds to either the causal past of the north observer or the causal future of the south observer. In terms of the Minkowski coordinates they are defined by:
\begin{equation}
    \eta = \frac{-1}{H^2 (X_0 +X_4)}
\end{equation}
\begin{equation}
    x_i = \frac{X_i}{H(X_0+X_4)} = -X_i \eta H
\end{equation}
with $X_0+X_4 > 0$. In the so defined planar patch we can introduce a {\it conformal time foliation} using hypersurfaces with fixed value of $\eta$ (see Fig.~\ref{fig:Fig5}). Clearly the Killing vector $\frac{\partial}{\partial \eta}$ is not globally defined in the planar patch. Hence the corresponding time evolution will lead to the standard creation of particles and will be characterized by a Bogolioubov transformation. We will introduce another natural foliation of the planar patch in terms of hypersurfaces having constant value of $\Lambda = k\eta H$. In Fig.~\ref{fig:Fig6} we present the Penrose cartoon showing how quantum fluctuations are stretched until they exit the horizon.

\subsection{Planar quantum Fisher information}

In order to define quantum fields in the planar patch we will start by introducing for a generic value of $|k|$ two Heisenberg algebras (ingoing and outgoing) \cite{MV1} with creation-annihilation operators $a_{|k|, \pm}$ and $a_{|k|, \pm}^{ \dagger}$ with
\begin{equation}\label{modes}
    [a_{|k|,+} a_{|k'|,+}^{+ \dagger}] = \delta(|k|-|k'|) = [a_{|k|,-}, a_{|k'|,-}^{\dagger} ]
\end{equation}
These operators define the mode expansion in planar coordinates $\eta,x$.
We can now formally define a {\it planar $\alpha$ vacua} as
\begin{equation}
    |\alpha> \equiv \e^{\int_{|k|} \mathcal{C} (\alpha) a_{|k|,+}^{\dagger} a_{|k|,-}^{ \dagger}} |0>
\end{equation}
The reason we refer to these states as formal $\alpha$ vacua is because the coefficient $\mathcal{C} (\alpha)$ entering into the exponent is independent of $|k|$. Let us now introduce the $|\eta\rangle$ vacua associated with a given conformal time as
\begin{equation}
    |\eta> \equiv \e^{\int_{|k|} \mathcal{C}_{dS} (\eta,|k|) a_{|k|,+}^{\dagger} a_{|k|,-}^{ \dagger}} |0> = \Pi |k, \eta>
\end{equation}
where $\mathcal{C}_{dS} (\eta,|k|)$ is determined by the condition
\begin{equation}
    a_{|k|,+}(\eta)|\eta\rangle = a_{|k|,-}(\eta)|\eta\rangle =0
\end{equation}
for all $|k|$ and with the algebra of creation annihilation operators at time $\eta$ being defined by the Bogolioubov transform
\begin{equation}
    a_{|k|,+}(\eta) = A(|k|,\eta) a_{|k|,+} + B(|k|,\eta) a_{|k|,-}^{\dagger}
\end{equation}
The coefficients $A$ and $B$ are determined by the de Sitter equations of motion and they have values \cite{MV1}
\begin{equation}
    A(|k|,\eta) = e^{-i\theta(|k|,\eta)} \cosh (r(|k|,\eta)
\end{equation}
and 
\begin{equation}
    B(|k|,\eta) = e^{i\theta(|k|,\eta)} e^{i\Phi(|k|,\eta)} \sinh (r(|k|,\eta))
\end{equation}
leading to
\begin{equation}
 \mathcal{C}_{dS} (\eta,|k|) = \tanh (r(|k|,\eta)) e^{2i\Phi(|k|,\eta)}
\end{equation}
with the squeezing parameter $r(|k|\eta)= - \sinh^{-1}(\frac{1}{2|k|\eta})$ and with the phase
given by
\begin{equation}\label{phase}
    \Phi(|k|\eta) = -\frac{\pi}{4}- \frac{1}{2} \tan^{-1}(\frac{1}{2|k|\eta})
\end{equation}
The global rotation $e^{-i\theta(|k|,\eta)}$ will not enter in the definition of the state $|\eta\rangle$ if we choose $|0\rangle$ rotationaly invariant.
Since the Bogolioubov coefficients depend on $|k|\eta$ we can define
for a fixed value of $\Lambda =|k|\eta H$ the $|\Lambda\rangle$ vacua as
\begin{equation}
    |\Lambda> = \e^{\int_{|k|} \mathcal{C}_{dS} (\Lambda)) a_{|k|,+}^{\dagger} a_{|k|,-}^{ \dagger}} |0>
\end{equation}
Note that the so defined state is a formal planar $\alpha$ vacua of the type defined above. By that we mean that the coefficient $\mathcal{C}_{dS} (\Lambda)$ is independent on the mode label $|k|$. Moreover the corresponding $\alpha(\Lambda)$ is characterized by
\begin{equation}
    {\rm Im} (\alpha(\Lambda)) = \Phi(\Lambda)= -\frac{\pi}{4}-\frac{1}{2} \tan^{-1}(\frac{H}{2\Lambda})
\end{equation}
Using the former ingredients we can define the {\it de Sitter renormalization group generator} as

\begin{equation}
    \frac{d |\Lambda>}{d \Lambda} = H^{{\rm RG}} |\Lambda>
\end{equation}
where $H^{\rm RG} (\Lambda) = \mathcal{C}_{dS} (\Lambda) a_{|k|,+}^{\dagger} a_{|k|,-}^{ \dagger} + c.c.$. Thus we can define the {\em Planar Quantum Fisher Information} as
\begin{equation}
    \boxed{\mathcal{F}_{Q} (\Lambda) \equiv Var[H^{\rm RG} (\Lambda)]}
\end{equation}
where $Var[H^{\rm RG} (\Lambda)] \equiv <\Lambda| (H^{\rm RG})^2 | \Lambda > - \left ( < \Lambda | H^{\rm RG} | \Lambda > \right )^2$.
Before entering into the numerical evaluation of this quantity let us stress some important differences between this quantity and the formal Quantum Fisher defined in (\ref{fisher}). First of all, the {\it Planar Quantum Fisher Information} defines a form of information that is accessible to a physical observer and is obtained on the basis of the statistical analysis of measurements performed on a causal patch. Secondly, the modes used in the expansion are the ones defined for the planar coordinates with well defined value of $|k|$. Third, the role of the formal $\alpha$ angle is now played by the physical momentum $\Lambda=|k|\eta H$. Finally we use $H^{\rm RG}$ to define the planar quantum Fisher or equivalently we parameterise with this Fisher information standard dilatations of the physical momentum.

\section{Numerical evaluation of the planar quantum Fisher Information}
Using the former set of ingredients we will proceed to evaluate
\begin{equation}\label{value}
    \mathcal{F}_{Q} (\Tilde{\Lambda}) = \sum_{N} |C(\Tilde{\Lambda})|^{2 N} \left (N \left .\frac{\partial \Phi}{\partial \Lambda} \right |_{{\scriptscriptstyle \Lambda = \Tilde{\Lambda}}} \right )^{2} - \left [ \sum_{N} |C(\Tilde{\Lambda})|^{2 N} N \left .\frac{\partial \Phi}{\partial \Lambda} \right |_{{\scriptscriptstyle \Lambda = \Tilde{\Lambda}}}    \right ]^{2}
\end{equation}

Note that expression (\ref{value}) is simply the variance of the generator defining changes in ${\rm Im}(\alpha(\Lambda))$. Before evaluating this quantity we will introduce a {\it pivot scale $|k_0|$} by $\Lambda=|k_0|\eta H$ and project the variation in $\Lambda$ at fixed $|k_0|$ by variation in $|\eta|$. This leads to
\begin{equation}
    F_{Q} (|k_{0}| \eta) \equiv |k_{0}|^{2} F_{Q} (\Lambda = |k_{0} \eta|)
\end{equation}
Using now the de Sitter data defined above, namely:
$C(\Lambda) = \tanh(r(\Lambda))$
with the squeezing parameter $r(\Lambda)= - \sinh^{-1}(\frac{H}{2\Lambda}$ and with the phase $\Phi(\Lambda)=-\frac{\pi}{4}- \frac{1}{2} \tan^{-1}(\frac{H}{2\Lambda})$ we will proceed
to the numerical evaluation of this quantity  introducing a cutoff $\cal{N}$ in the sum. The result is given by
\begin{equation}
    F_{Q} (|k_{0}|, \eta, {\cal{N}}) = \frac{1}{\eta^2} \frac{1}{(|k_{0}| \eta)^{6-2 \alpha_{F}}}.
\end{equation}
with the {\it quantum tilt} $\alpha_F(|k_0|\eta;{\cal{N}})$. Before proceeding further let us make some general comments. First of all the {\it quantum tilt} is fully determined by the numerical evaluation as a function on $|k_0|\eta$ and the cutoff $\cal{N}$ of the sum in (\ref{value}). In Fig.~\ref{fig:tilt} we plot the numerical value of $\alpha_F$ evaluated using $\cal{N}$ of order $\mathcal{O}(10^9)$. Defining
\begin{equation}
\mathcal{F}_{Q} (|k_{0}|, \eta, {\cal{N}})  = (|k_{0}| \eta)^{6} F_{Q} = \frac{1}{\eta^2} \frac{1}{(|k_{0}| \eta)^{-2 \alpha_{F}}}
\end{equation}
we extract the {\it anomalous} scale dependence of the planar quantum Fisher information that we can encode in the following renormalization group equation:
\begin{equation}
    \boxed{|k_{0}| \frac{d}{d |k_{0}|} \mathcal{F}_{Q} (|k_{0}|, \eta, {\cal{N}}) = 2 \alpha_{F}(|k_0|\eta;{\cal{N}}) \mathcal{F}_{Q} (|k_{0}|, \eta, {\cal{N}})}
\end{equation}
\section{The Quantum Tilt $\alpha_{F}$ and $\mathcal{N}$ sensitivity}
\begin{figure}
    \centering
    \includegraphics[width=0.7\columnwidth]{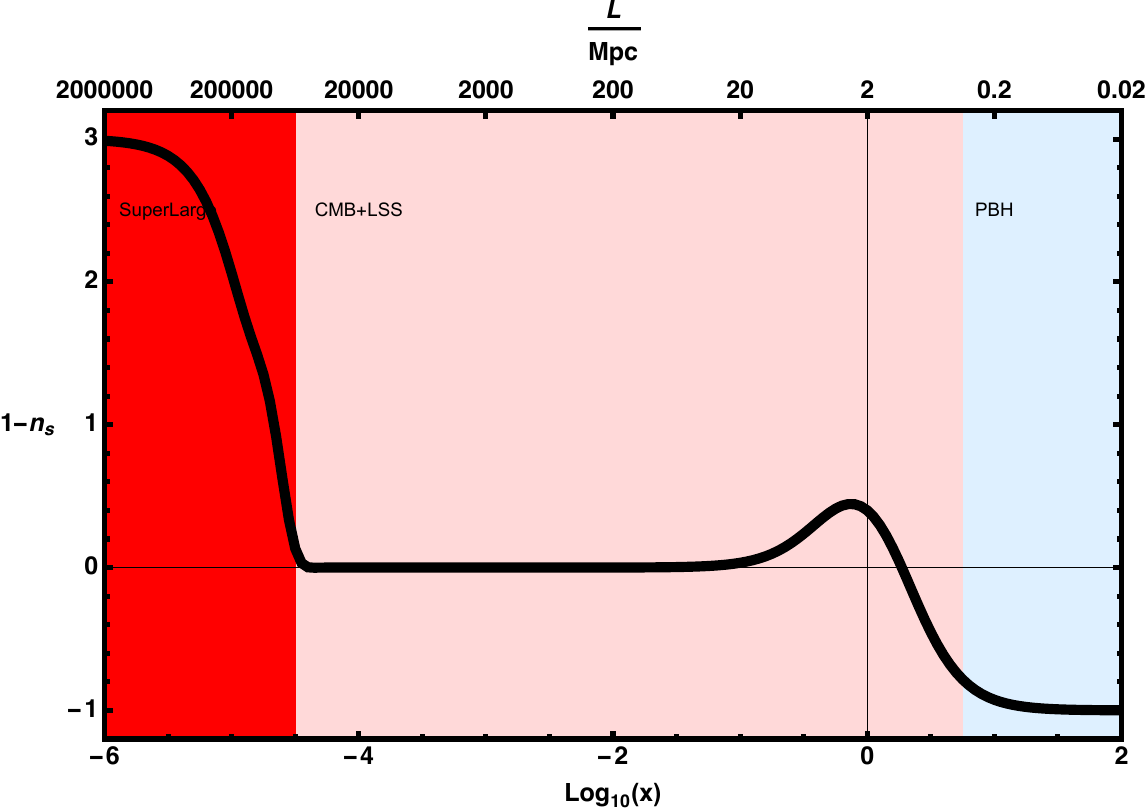}
    \caption{The quantum cosmological tilt as a function of energy scale as determined by the quantum Fisher. For values of $x=|k \eta| < 0$ the end of inflation is determined by the IR cutoff on the number of particles $\mathcal{N}$. In this plot, for illustrative purposes, we set $\mathcal{N}=10^{9}$. The different shaded regions correspond to different scales. The blue region is where the dark matter forms and the spectral index is blue.}
    \label{fig:tilt}
\end{figure}

The numerical result for $\alpha_F(|k_0|\eta; {\cal{N}})$ for ${\cal{N}}$ of the order of $10^9$ is plotted in Fig.~\ref{fig:tilt}. Before discussing the sensibility of this numerical result on the cutoff ${\cal{N}}$ let us comment on some significant aspects of this plot. The first thing we observe is the existence of an almost flat region where the value of $\alpha_F$ is almost constant and very small. As we will discuss later, this region is the one we can associate with large scales of the order of the CMB scale. When we move to the left, we move into much larger scales or smaller momentum. The first curious fact we find when we move into this IR regime is the sudden change of the value of $\alpha_F$. What is the meaning of this jump at super larger scales? Super larger scales corresponding to deep IR values of the momentum will normally be interpreted as providing us information about the initial conditions of inflation. In other words, in a generic scenario based on an inflaton potential, these superlarge scales depend on the characterization of the initial conditions of inflation since these scales exit the horizon very early. If we take seriously the numerical plot the point in the IR where $\alpha_F$ jumps can be interpreted as the larger scales that exit the primordial horizon. However in order to take seriously this IR region a crucial necessary previous ingredient is to know the sensibility of the plot in this region on the numerical cutoff ${\cal{N}}$. 

\begin{figure}
    \centering
    \includegraphics[width=0.7\columnwidth]{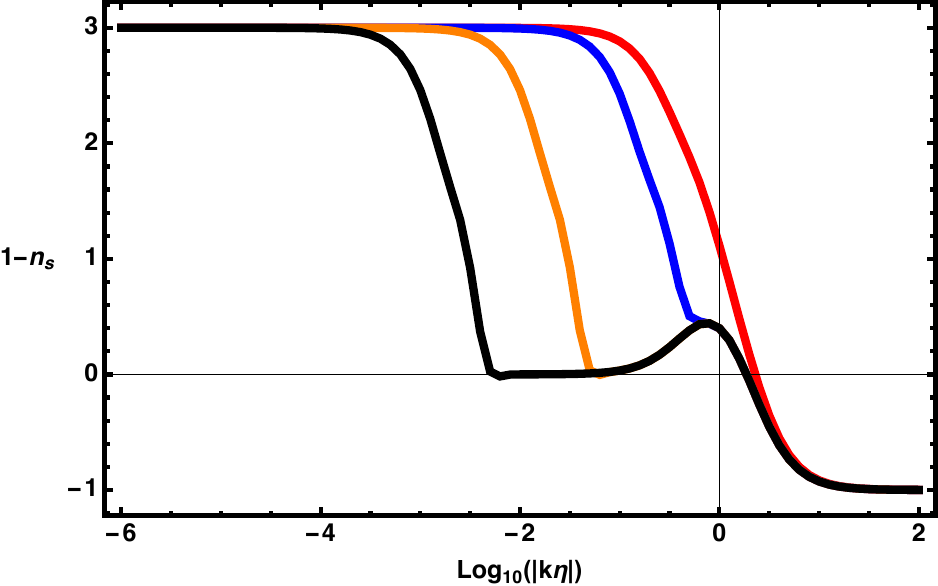}
    \caption{The effect of the number of pair production $\mathcal{N}$ on the cosmological tilt $1-n_s$. The red, blue, orange and black curves are for $\mathcal{N} = 1, 10, 1000, 10^{5}$ respectively. The nearly scale invariant tilt is produced by the increase in $\mathcal{N}$. See text for more details on how to set $\mathcal{N}$ and end inflation.}
    \label{fig:tilt_n}
\end{figure}

As shown in Fig.~\ref{fig:tilt_n} the sensibility relative to the value of ${\cal{N}}$ is very important in the IR regime by contrast to what happens in the UV regime that is very stable with respect to changes of ${\cal{N}}$. What we observe numerically is that when we send ${\cal{N}}$ to extreme large values the jump in the IR takes place around $\frac{1}{\sqrt{{\cal{N}}}}$. A potentially interesting conjecture could be to put an upper bound on the value of ${\cal{N}}$ based on the standard GH entropy for the de Sitter space. The logic underlying this conjecture is to impose that the entropy of ${\cal{N}}$ entangled pairs contributing to the quantum planar Fisher information is at most of the order of GH entropy. This is just a very qualitative conjecture that requires more input and further analysis.

Concerning the UV part of the plot we observe an interesting change of regime with $\alpha_F$ becoming negative at small scales or large values of the momentum. We will make below some general comments later on the physical meaning of this change of regime.

\section{Model Independent Derivation of Spectral Index}
At this point, we arrive to the central core of this article, namely the attempt to provide a model independent derivation of the cosmological spectral index we measure using CMB data. A priori, this attempt seems futile for several reasons. As already discussed, this is based on suspecting some form of universality of the spectral index in the same way we are used to think in universality when treating with critical phenomena. In the particular case of the cosmological spectral index, the bold claim about model independence seems to indicate that the type of information encoded in the spectral index, that becomes observationally manifest in the CMB, is already secretly contained in the quantum statistical properties of pure de Sitter. The obvious reason for which this claim sounds a priori bizarre is that the spectral index we observe, at CMB scales, seems to be directly related with the number of e-foldings lasting between the moment this scale leaves the primordial de Sitter horizon and the end of inflation. However, in pure de Sitter there is not any classical notion of graceful exit. Hence, the standard view that relates the spectral index with a classical graceful exit seems to immediately rule out our model independence expectations. Moreover, what we actually observe depends on working with a particular large scale regime, the CMB scale, where the simplified Sachs-Wolf representation of temperature fluctuations is working. So, why pure de Sitter should know about this peculiar features underlying the actual observations? The answer to this question is summarized in the following steps:

\begin{enumerate}
    \item The first step is based on the observation that a planar quantum Fisher information for pure de Sitter immediately allows us to define an intrinsic energy uncertainty, statistical in origin, in the dispersion relation of the modes defining the Mukhanov-Sasaki gauge invariant variable for pure dS. This statistical uncertainty in the time dependent frequency carries the key information on what we have denoted as the quantum tilt $\alpha_F$, that as stressed several times, is a pure de Sitter information. Moreover this statistical uncertainty in the frequency depends non trivially on the conformal time $\eta$ and the pivot scale $|k_0|$.
    \item The second step identifies the standard contribution to the mode frequency due to modifying de Sitter in the form of a generic qde Sitter that we parametrize, in terms of t $\delta_{qdS}$. As discussed before, this qde Sitter parametrization in terms of $\delta_{qdS}$, of a generic one field inflaton potential, contains information about the whole set of the corresponding slow roll parameters in the region of large scales.
    \item In the final step we identified the statistical Fisher uncertainty of the mode frequency with the qde Sitter expression. In other words, we ask ourselves what $\delta_{qdS}$ accounts for exactly the quantum Fisher statistical uncertainty. As we will show this condition uniquely sets the value of $\delta_{qdS}$ as well as the value of $|k_0|\eta$ for which this quasi de Sitter effective description is working. 
    \item The former solution is identified with the CMB large scale where our quasi de Sitter parametrization is obviously working. 
\end{enumerate}

Although the results we will present are in perfect agreement with the observations, it is obvious that a priori we do not  have any solid reason to expect such agreement. It could be perfectly possible that the statistical Fisher uncertainty leads to a smaller value of the spectral index. That this is not the case seems to have a deep meaning. Quantum Fisher defines an upper bound, 
so if this procedure gives a smaller deviation, relative to what we observe, from scale invariance, it will indicate that the primordial information transfer to the moment of reentering is bigger than the maximal pure de Sitter information we can have about the scale invariant features of what exits the primordial horizon during the inflationary era. Note that we are only trying to catch scale dependence but not the concrete values of the amplitudes.

Let us now go through all these steps in detail. First of all, let us focus on the phases of the state $|\Lambda\rangle$ given by (\ref{phase}) and let us introduce a {\it formally conjugated} energy by
\begin{equation}
    \Phi(|k_0|\eta) \equiv \eta \mathcal{E} (|k_{0}| \eta)
\end{equation}
The quantum Fisher variance $\langle \Delta \mathcal{E}^2 \rangle$ of $\mathcal{E} (|k_{0}| \eta)$ for each mode is determined by the quantum Fisher, hence we get
\begin{equation}
     \delta_{F} \mathcal{E} (|k_{0}| \eta) = \frac{1}{2}\sqrt{\mathcal{F_{Q}}} =  \frac{1}{2\eta} (|k_{0}| \eta)^{\alpha_{F} (|k_{0}| \eta) }
\end{equation}
where we introduce the factor $\frac{1}{2}$ to account for the energy of one of the two modes we have introduced in (\ref{modes}). Note that to define the uncertainty of $\mathcal{E} (|k_{0}| \eta)$ we use the square root of the quantum Fisher information. The reason for this is that we are interested in extracting the Fisher uncertainty for the frequency square of the modes at first order. This will be defined as
\begin{equation}
\omega_0 \delta_{F} \mathcal{E} (|k_{0}| \eta)
\end{equation}
with $\omega_0=|k_0|$. Finally, in order to define the physical energy uncertainty we use the pure de Sitter metric leading to
\begin{equation}
    \Delta_{F} {E}^{2} (|k_{0}| \eta) \equiv \frac{|k_0| \delta_F \mathcal{E}(|k_0|\eta)}{a_{dS}^{2}} = \frac{1}{2} |k_{0}| \eta H^{2} (|k_0| \eta)^{\alpha_{F} (|k_{0}| \eta) }
\end{equation}
This is the quantity that we want to compare with the quasi de Sitter contribution. In order to define this quantity we work in a generic qde Sitter parametrized by a generic $\delta_{qdS}$. The qde Sitter effect on frequencies is:
\begin{equation}
    w_{qdS}^{2} - w_{dS}^{2} = \frac{3 \delta_{qdS} + \delta_{qdS}^2}{\eta^2}
    \label{eq:fisher}
\end{equation}
Now to define the qde Sitter contribution to the physical energy we use the  metric ${\tilde a}_{qdS}\equiv \frac{a_{qdS}}{\sqrt{\epsilon_0}}$ with $a_{qdS}$ defined in (\ref{metric2})\footnote{Note that locally the physical metric is defined by $\frac{a_{qdS}}{\sqrt{\epsilon_0}}$.} . This leads to
\begin{equation}
    \Delta_{qdS} E^{2} (|k_{0}| \eta) = \frac{ (3 \delta + \delta^2)}{\eta^2 {\tilde a_{qdS}}^2} =  (3 \delta + \delta^2) H^{2} (|k_{0} \eta)^{2 \delta}
    \label{eq:qds}
\end{equation}

\begin{figure}
    \centering
    \includegraphics[width=0.7\columnwidth]{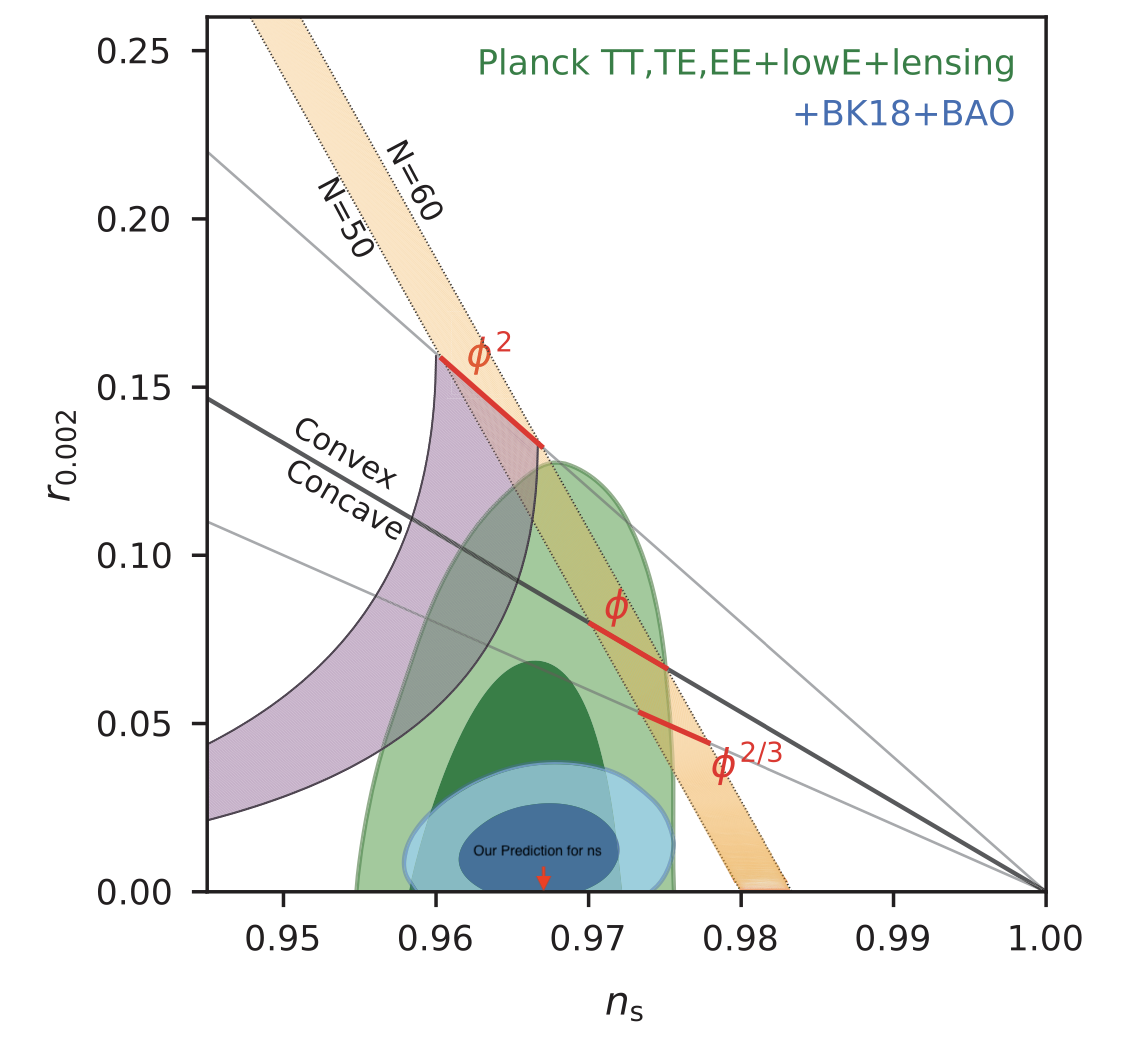}
    \caption{Figure adapted from the latest BICEP/Keck Array results \cite{BICEP}. The blue shaded region is their stringent constraint on $n_s$. The red arrow depicts our prediction.}
    \label{fig:biceps}
\end{figure}

The final step is to look for the values of $\delta_{qdS}$ and $|k_0||\eta|$ such that
\begin{equation}
    \Delta_{F} \E^{2} (|k_{0}| \eta) = \Delta_{qdS} (E^{2} (|k_{0}| \eta)
\end{equation}
This leads to two simple equations
\begin{eqnarray}\label{index}
       2 (3 \delta + \delta^{2}) & = & |k_{0}| \eta \\ 
       \alpha_{F} (|k_{0}| \eta) & = & 2 \delta
\end{eqnarray}
that we can solve using as data the pure de Sitter quantum tilt $\alpha_{F} (|k_{0}| \eta)$. Note that both equations combine into an {\it almost fixed point} equation
\begin{equation}
    \alpha_{F} (6 \delta + 2 \delta^2) = 2 \delta
\end{equation}
with unique solution $2\delta_{qdS} = 0.0328$ and $|k_{0}| |\eta| = 0.1$. Using now the qde Sitter parameterization of $(1-n_s)$ as $2\delta_{qdS}$ we get the desired prediction on the spectral index as well as the corresponding value of $|k_0||\eta|$ (see Fig.~\ref{fig:biceps} for the recent experimental results on the allowed values of the spectral index ).

\subsection{Confronting with Observational Data}

Recall first how the value of $n_s$ is obtained experimentally. This is done in a fully model dependent way. First a parametric shape of the primordial power spectrum is adopted. In the simplest case a power law with power $n_s$. Then using this primordial power spectrum, a Boltzmann hierarchy and a cosmological model, a prediction is made for the observed power spectrum of the CMB and LSS. This is fit to the data using a Bayesian method and the best value of $n_s$ extracted (see~\cite{Planck18}). One can use a more sophisticated model and add running, i.e., $n_s(k)$ depends on scale. Or can go a step ahead and fit the least parametric possible way by using spline reconstruction (see~\cite{Sealfon} for the original idea and~\cite{Ravenni} for latest results). Obviously, as the method becomes more model independent, the constraints on $n_s$ become less restrictive \cite{Planck18,Ravenni}. Note that the above method to extract information on the value of $n_s$ is independent of the inflation model. Using the number of e-folds for  a given scale between exit and re-entering the horizon to constrain $n_s$ is fully specific to the particular model of inflation considered.

In order to confront our result with observational data let us revisit how the prediction in the  Starobinsky model works. Once we fix the scale where we perform observations by $k_{{\rm CMB}}$, the spectral index at that scale is given by
\begin{equation}\label{slava}
    (1-n_s)_{\rm CMB} = \frac{2}{\ln(k_{{\rm CMB}}\eta_{{\rm end}})}
\end{equation}
where $\eta_{\rm end}$ is the conformal time at the end of inflation. This produces the right experimental result if (and only if) the number of e-foldings between the moment the scale $k_{\rm CMB}$ exits the horizon and the end of inflation is between 50 and 60~\footnote{In this model the value of the scalar tilt is fine tuned to the number of e-folds, so if the tilt was to be measured with, say, four significant figures, the number of e-foldings would have to be fine-tuned to one part in $10^{7}$ }. In the language of $R^2$ gravity the former expression reduces to
\begin{equation}\label{one}
    (1-n_s) = \frac{4}{3} e^{-\sqrt{2/3} \phi_{i}}
\end{equation}
for $\phi$ the canonically normalized inflaton field and with the value $\phi_{i}$ in (\ref{one})
the one associated to the time at which the CMB scale exits the horizon. This formula is approximate and assumes that the value of $\phi$ at the end of inflation is much smaller than $\phi_{i}$ \footnote{ Using $\phi=\sqrt{3/2}\ln (1+\frac{R}{4M^2M_P^2})$ one recovers, in this approximation, the relation  $(1-n_s) = \frac{2M^2}{3H^2}$.}.

In the above discussion on the model independent derivation of the spectral index we observed that  Quantum Fisher fits the quasi de Sitter result in what we can denote the {\it fixed point regime}. More specifically, given $\alpha_{F}(x)$ where we denote the argument of $\alpha_F$ as $x$ the {\it fixed point regime} is characterized by
\begin{equation}
    \alpha_F(6 x) \sim 2x
\end{equation}
Thus the first ingredient we use to confront our results with the experiment is to associate the {\it fixed point regime} of $\alpha_F$ with the CMB regime of large scales where (\ref{slava}) and (\ref{one}) are working i.e.
\begin{center}
  {\it fixed point regime of} $\alpha_F$  $\rightarrow$ {\it CMB large scales}
\end{center}

Let us now come back to (\ref{slava}). How to extended this formula to scales smaller or larger than CMB? This is a crucial question in order to contrast experimentally the whole picture beyond the current experimentally accessible scales. For us this question translates into: How to interpret $\alpha_F$ beyond the fixed point regime? Here we can only offer a conjecture that leads to  well defined experimental predictions. Namely, setting the CMB scale by the fixed point condition corresponding to $x\equiv|k_0||\eta|\sim 0.1$ we suggest, working in a linear approximation, that $(1-n_s)$ at scales $ak_{\rm CMB}$ for $a$ a numerical proportionality factor is given by
\begin{equation}\label{map}
    (1-n_s)(a k_{\rm CMB}) = \alpha_F(x=\frac{a}{10})
    \end{equation}
 Using the qualitative relation (\ref{map}) we can extract some qualitative information on the region out of the CMB {\it fixed point regime}. Although we will not cover in this article this phenomenology, let us just mention two interesting facts. On one side we can evaluate the running of the spectral index in the region of large scales that turns out to be $-0.0019$ which is in perfect agreement with the Planck18 bounds. The typical scale at which we expect a change of regime from red tilted to blue tilted is of the order of $1 {\rm Mpc}$. 
 
 We must stress once more the logic underlying these results. The expected spectral index of the primordial power spectrum for values of the momentum larger than the CMB momentum is assumed to be determined universally by the quantum tilt $\alpha_F$. However, we  are aware that in regions of large momentum where we cannot use Sachs-Wolfe approach what we observe can differ  from our prediction due to the departure from the  Sachs-Wolfe relation. This is a subtle issue to study so the former predictions are just qualitative.

\subsection{A comment on tensor modes}
What we have shown in the previous sections is how to use the quantum Fisher information to fix the value of $\delta$ defining the scaling of $\frac{z^{''}}{z}$. This uniquely fixes the spectral index for scalar perturbations. However as discussed in section I-E the power spectrum for the tensor modes is not fully determined by the value of $\delta$ and requires extra information on $\epsilon$. This extra information is {\it model dependent} and cannot be extracted using only the quantum Fisher information. This is phenomenologically very interesting. Indeed quantum Fisher fixes $(1-n_s)$ in a model independent way but {\it not} the tensor spectral index that we have not yet experimentally observed. The technical reason for that is easy to understand. The power spectrum for the tensor modes requires to disentangle the correction to the frequency into two pieces the one associated with $\epsilon$ and the one determined by $\eta$. However the quantum Fisher correction gives us the global value of $\delta$ and is not allowing us to disentangle both contributions in a model independent way. In other words the tensor modes encode primordial information that goes beyond the quantum Fisher information setting the scalar spectral index. Actually the  primordial dS planar quantum Fisher information is consistent with a tensor to scalar ratio as small as we wish. Thus even if tensor modes are not observed the quantum Fisher prediction for $(1-n_s)$ presented here will be unaffected. 
\section{Concluding Remarks}
The main conceptual lesson we learnt from our analysis is that the anomalous scale dependence of the primordial spectrum of curvature scalar fluctuations is fully determined by the scale transformations of the pure de Sitter quantum Fisher information reduced to the planar patch. In more technical words: the anomalous scale invariance of the primordial spectrum is encoded in the variance, for a planar observer Hilbert space, of the generator of dilations of physical momentum. This primordial quantum information on primordial de Sitter captures what a graceful exit quasi de Sitter mechanism makes observable after reentering. We insist that this only affects the scale dependence and not the amplitudes. As already mentioned, the predictions on scales different from the CMB scale depend on how the transfer functions defining through convolution the observable spectrum behave in those scales.

\begin{acknowledgments}
We thank the anonymous referee for a very thoughtful and positive report. The work of CG is supported by grants SEV-2016-0597, FPA2015-65480-P and PGC2018-095976-B-C21. The work of RJ is partially supported by  project PGC2018-098866-B-I00 MCIN/AEI/10.13039/501100011033 y FEDER ``Una manera de hacer Europa'', and the  ``Center of Excellence Maria de Maeztu 2020-2023'' award to the ICCUB (CEX2019- 000918-M funded by MCIN/AEI/10.13039/501100011033).
\end{acknowledgments}

\appendix*
\section{Quantum Fisher Imformation}
\label{app:fisher}

In this appendix we introduce the basics of Quantum Fisher Information.  Consider $P(x)$ the probability distribution of $x$ for an external parameter $\theta$. Then the classical Fisher information $F$ is:

\begin{equation}
F(\theta) = \int ds P(x; \theta) \left ( \frac{d \ln P(x; \theta)}{d \theta} \right )^{2} = \int dx \left ( \frac{dP}{d \theta} \right )^{2} P^{-1} 
\end{equation}

This provides the statistical information about the external parameter $\theta$. The minimum error on the parameter can be no smaller than $F^{-1/2}$. Let us illustrate this with one example. Imagine a dice that is loaded in such a way that the loading depends purely on the temperature of the room where the dice is rolled. The classical Fisher is simply the sum over all trial where the dice is rolled in rooms at different temperatures.

Let us now move onto the quantum case. Let us write down a pure quantum state depending on the external parameter $\theta$ as

\begin{equation}
| \psi (\theta) \rangle = \sum c_{n} (\theta) e^{i \phi_{n} (\theta)} | n \rangle 
\end{equation}

with $P(n ; \theta) = |c_{n} (\theta) |^2$. The Quantum Fisher Information is the same as the classical but now we define a pure quantum piece using the phases of the probability amplitude, such that

\begin{equation}
F_{Q} (\theta) = \sum P(n; \theta) \left ( \frac{\partial \phi_{n}}{\partial \theta} \right )^{2} - \left (\sum P(n; \theta) \frac{\partial \phi_{n}}{\partial \theta} \right )^{2} \equiv {\rm Var} \left [ \left ( \frac{\partial \phi_{n}}{\partial \theta}  \right )^{2} \right ]
\end{equation}

Let us illustrate this with an example. Take  $e^{i \phi_{n} (\theta)} = e^{i E_{n} t}$ then $F_{Q} = \langle E^{2} \rangle - \langle E \rangle ^{2}$. In general we will have that
\begin{equation}
F_{Q} = \langle H_{\theta}^2 \rangle - \langle H_{\theta} \rangle^2
\end{equation}

Now it is very easy to see that for the dS case, the Quantum Fisher information is simply
\begin{equation}
F_{Q} =  \langle E^{2} (k \eta) \rangle - \langle E(k \eta) \rangle ^{2}
\end{equation}
with $E$ the oscillator energy with comoving momentum defined at fixed $k$ at time $\eta$.

\end{document}